\documentclass{resonance}
\usepackage{amsfonts}
\usepackage{amsmath}
\usepackage{amssymb}
\usepackage{rotating}
\usepackage{epsfig}  
\usepackage{graphics} 
\usepackage{graphicx} 
\usepackage{txfonts}
\usepackage{marginnote}
\usepackage{array}
\usepackage{geometry}
\usepackage{color}      
\definecolor{light}{rgb}{0.8,0.5,0.5}
\definecolor{cblue}{rgb}{0.9,0.9,1.0}
\definecolor{darkblue}{rgb}{0.1,0.1,0.6}
\definecolor{darkred}{rgb}{0.6,0.1,0.1}
\usepackage{cleveref}
\crefformat{section}{\S#2#1#3}
\crefformat{subsection}{\S#2#1#3}
\crefformat{subsubsection}{\S#2#1#3}
\crefrangeformat{section}{\S\S#3#1#4 to~#5#2#6}
\crefmultiformat{section}{\S\S#2#1#3}{ and~#2#1#3}{, #2#1#3}{ and~#2#1#3}
\newcommand{\bed}{\begin{displaymath}}
\newcommand{\eed}{\end{displaymath}}
\newcommand{\bei}{\begin{itemize}}
\newcommand{\eei}{\end{itemize}}
\newcommand{\bef}{\begin{figure}}
\newcommand{\eef}{\end{figure}}
\newcommand{\ben}{\begin{enumerate}}
\newcommand{\een}{\end{enumerate}}
\newcommand{\beq}{\begin{equation}}
\newcommand{\eeq}{\end{equation}}
\newcommand{\ber}{\begin{eqnarray}}
\newcommand{\eer}{\end{eqnarray}}

\newcounter{attnctr} 

\begin{document}

\title{The Sounds of Music : Science of Musical Scales}
\secondTitle{II : Western Classical}
\author{Sushan Konar}

\maketitle
\authorIntro{\includegraphics[width=2.5cm]{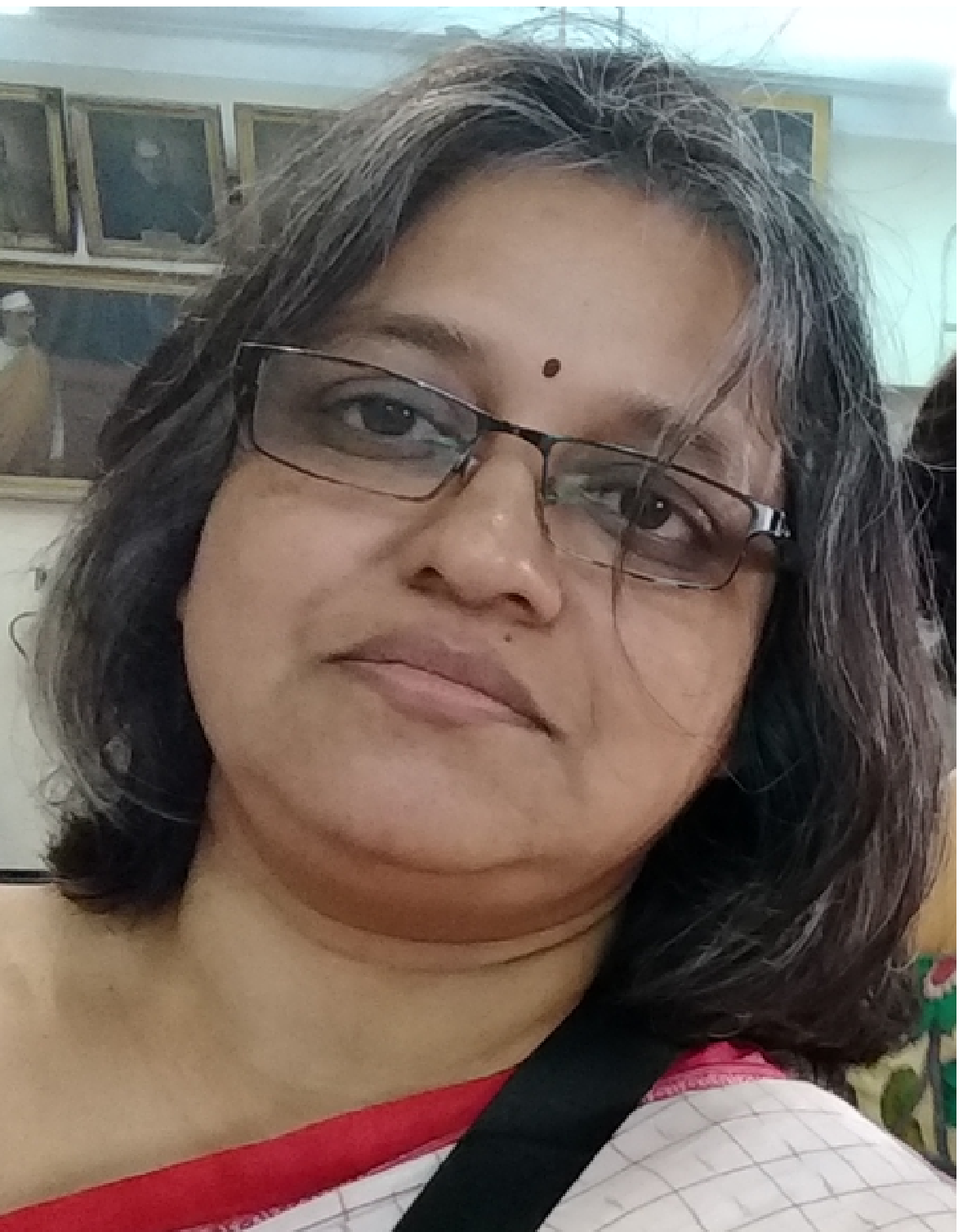}\\
  Sushan Konar  works  on  stellar  compact objects. She also writes popular
  science articles and maintains a weekly astrophysics-related blog called
  `{\em Monday Musings}'.}
\begin{abstract}
  A set of  basic notes, or `scale', forms the  basis of music. Scales
  are specific to  specific genre of music. In this  second article of
  the series we  explore the development of  various scales associated
  with western classical  music, arguably the most influential genre of
  music of the present time.
\end{abstract}
\monthyear{2018}
\artNature{GENERAL  ARTICLE}

%%====================================================================%%

\section*{Introduction}
\label{s-intro}
Just like an alphabet is the basis of a language, a set of basic notes
known as a {\bf \em scale} forms the basis of music.  The term `scale'
comes from the Latin word `scala' meaning `ladder'.  Thus a scale is a
ladder or a set of notes  ordered according to their frequency or {\em
  pitch}. A scale  ordered by increasing pitch is  an ascending scale,
and a scale ordered by decreasing pitch is a descending scale. Scales,
in general, may or may not contain the same pitches while ascending as
well as  when descending.  Moreover, even  all of the scale  steps may
not be equal. Indeed, most of  the traditional music scales began with
unequal  scale steps.   Due to  the principle  of octave  equivalence,
scales are generally  considered to span a single  octave, with higher
or  lower  octaves  simply  repeating the  pattern.  A  musical  scale
represents a  division of the octave  space into a specific  number of
scale steps,  a scale step  being the interval between  two successive
notes of the scale.


There exist  a number  of scales  depending on  the number  of primary
notes (or fundamental frequencies) available per octave.  However, the
traditional  `diatonic'  (in   Greek,  `diatonic'  means  `progressing
through tones') scale used in Western classical music is heptatonic (7
notes  per octave).  Evidently,  in a  heptatonic  scale, doubling  of
frequency  requires going  up  by  8 notes.  The  name  octave, for  a
non-repetitive frequency span, derives from  this. In the following we
shall discuss  the origin of  the heptatonic scale, starting  from the
{\bf  \em Pythagorean  scale}  (one of  the  first theoretical  tuning
structures known)  and its  subsequent transformation into  the modern
{\bf \em Equal Tempered Scale} (ETS),  with 12 notes per octave, where
all  the notes  are equidistant  (in a  logarithmic sense)  from their
neighbouring ones.   In this  section we shall  try to  understand the
mathematical    logic     behind    these    scales.

\section{Pythagorean Scale} 
\label{s-pytha}
Pythagoras  ({\em   circa}  500~BC),   the  Greek   mathematician  and
philosopher, used  a {\em mono-chord}  to study - a)  the relationship
between the  string length and  the note  produced when it  is plucked
(see  \S1.1,  article-I), and  b)  the  phenomenon of  consonance  and
dissonance.  A mono-chord, as seen  in Fig.[\ref{f-mono}], is a single
stringed  instrument with  a movable  bridge, dividing  the string  of
length $L$ into two segments, $l_1$  and $l_2 \, (=L-l_1)$.  Thus, the
two string segments can have any desired ratio, $x = l_1/l_2$.

\begin{figure}[h]
\caption{A Pythagorean Mono-chord  - A single string  (chord) of length
  $L$ is divided by a movable bridge into two segments of length $l_1$
  and $l_2$ such that $l_1 + l_2 = L$.}
\label{f-mono}
%
%\begin{center}
\centering\includegraphics[width=12.5cm]{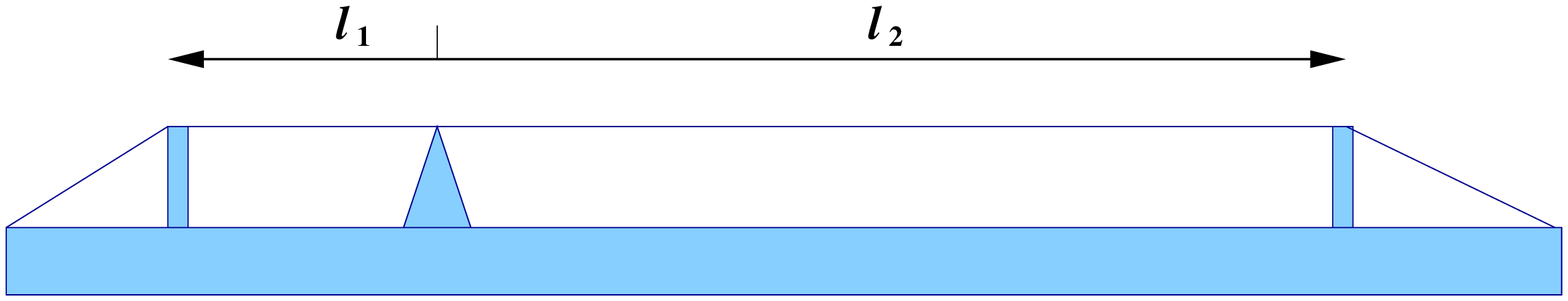}
%\end{center}
%
\eef
It is evident that the string tension $T$ and the mass per unit string
length  $\mu$ are  the same  in both  the string  segments.  When  the
mono-chord  is played,  both  string  segments vibrate  simultaneously.
Now, according to  Eq.[8] of article-I, when the string  is plucked the
frequencies  of  the  fundamental  vibrations  generated  in  the  two
segments are,
\beq
\nu_1 = \frac{1}{2l_1} \sqrt{\frac{T}{\mu}}, \; \;
\nu_2 = \frac{1}{2l_2} \sqrt{\frac{T}{\mu}} \,. 
\eeq
Therefore, the ratio of these vibrations would be -
\beq
\frac{\nu_2}{\nu_1} = \frac{l_1}{l_2} = x \,.
\eeq
It has  been found that for  $x = 1, 2,  3/2, 2, 4/3, 5/4,  6/5..$ the
sound combination  appears pleasing,  in other words  {\em consonance}
occurs.  These  ratios actually  correspond to  the -  unison, octave,
fifth, fourth,  major third,  minor third (some  of the  most pleasing
combinations) used  in the western classical  music. Dissonance occurs
when $x$ is such  that it can not be expressed as  a ratio between two
small integers.

It is believed that a tetra-chord (a four-stringed instrument) was used
to generate the firs set of  tuning (or fixing the scale).  Consider a
tetra-chord with 4 strings of length $l_1, \, l_2, \, l_3, \, l_4$ with
corresponding fundamental  frequencies $\nu_1,  \, \nu_2, \,  \nu_3, \,
\nu_4$.  The two  outer strings differed in length by  a factor of 4/3
and  so the  frequencies differed  by a  factor of  4/3 as  well.  The
middle two strings were tuned so that the ratio of the first to second
strings (the first  interval) was approximately the same  as the ratio
of the second to the third string. Therefore, we have
\ber
l_1 : l_4 &=& 4/3 \,, \\
l_1 : l_2 &=& l_2 : l_3 \,.
\eer
Assuming all the  strings to have the same tension  ($T$) and the same
mass per unit length ($\mu$) (using Eq.[8], article-I) we obtain,
\ber
\nu_1 : \nu_4 &=& 3/4 \,, \\
\nu_1 : \nu_2 &=& \nu_2 : \nu_3 \,, \\
\Rightarrow \nu_3 : \nu_4 &=& (\nu_2 : \nu_1)^2 \times \frac{4}{3} \,.
\eer
When $\nu_1 : \nu_2 = 9/8$, we obtain $\nu_3 : \nu_4 = 265/243$.  This
choice is known  as the {\em diatonic} tuning and  is the precursor of
the modern western scale. Now let us try to understand how this choice
comes about.

Pythagoras is supposed to have  used the two fundamental principles of
psycho-acoustics -  the octave equivalence,  and the consonance  of the
fifth. Using these two principles we can build up a set of basic notes
within an octave  which would sound `consonant'  when played together.
Starting with a base note (say, of frequency $\nu$) we obtain the next
one by multiplying it by 3/2 (a `perfect fifth up' according to musical
terminology), then obtain  the next one by multiplying  the second one
by 3/2. So we now have three  consonant frequencies - $\nu_1 = \nu, \,
\nu_2 = 3\nu/2, \, \nu_3 = 9\nu/4$.  But $\nu_3 > 2 \nu_1$. This means
that we have gone beyond the  first octave based at $\nu_1$. To obtain
the note corresponding  to the starting octave we divide  $\nu_3$ by 2
to obtain the new $\nu_3 = 9\nu/8$ (because of octave equivalence). We
can now continue  the process of multiplying by 3/2  and dropping down
to the correct octave to obtain the following set of notes -
\beq
\nu, \, \frac{3\nu}{2}, \, \frac{9\nu}{4} \sim 
\frac{9\nu}{8}, \, \frac{27\nu}{16}, \, \frac{81\nu}{32} \sim 
\frac{81\nu}{64}, \, \frac{243\nu}{128}, \, \frac{729\nu}{256} \sim 
\frac{729\nu}{512}...
\eeq
Collecting the frequencies within the first octave and ordering them
in an ascending order, we obtain -  
\beq
\nu, \,
\frac{9\nu}{8}, \,
\frac{81\nu}{64}, \,
\frac{729\nu}{512}, \,
\frac{3\nu}{2}, \,
\frac{27\nu}{16}, \, 
\frac{243\nu}{128}, \,
2 \nu.
\eeq
where the  note at double  the frequency has been  added artificially.
Note that the fifth term in this  series is exactly 3/2 times the base
frequency  and therein  lies  the explanation  for  the name  `perfect
fifth'\footnote{In Indian  music too, the  fifth note is  special.  We
  shall  see  later how  the  note  called {\em  pancham}  (literally
  meaning  the  `fifth')  has  been given  special  status  in  Indian
  tradition.}  The ratios of the  consecutive notes then appear in the
following order -
\beq
\frac{9}{8} :
\frac{9}{8} : 
\frac{9}{8} : 
\frac{256}{243} :
\frac{9}{8} : 
\frac{9}{8} :
\frac{256}{243} \,.
\label{e-octave}
\eeq
Therefore, the adjacent  notes differ either by a ratio  of either 9/8
(known as a Pythagorean  `tone') or by a ratio of  256/243 (known as a
Pythagorean   `semi-tone').    As   $(256/243)^2  \simeq   9/8$,   the
nomenclature is  more or less  justified. Pythagoras supposed  to have
used  these   two  ratios  or   musical  intervals  in   his  diatonic
tetra-chord. Therefore, Pythagorean tuning  of a tetra-chord (with respect
to the base note $\nu$) would be the following -
\beq
\frac{4 \nu}{3}, \, \frac{32 \nu}{27}, \, \frac{256 \nu}{243}, \, \nu.
\eeq

It is to be noted that if the pattern of scale-steps given by Eq.[\ref{e-octave}]
is repeated successively, like the following -  
\beq
...
\frac{256}{243},
\frac{9}{8},
\frac{9}{8},
\frac{256}{243},
\frac{9}{8},
\frac{9}{8},
\frac{9}{8},
\frac{256}{243},
\frac{9}{8},
\frac{9}{8},
\frac{256}{243},
\frac{9}{8},
\frac{9}{8},
\frac{9}{8},
\frac{256}{243}
... \,,
\eeq
then at every eighth step  the frequency (approximately) doubles. This
is the  reason for  the term  {\em octave} to  be associated  with the
doubling of a frequency. If `T'  (for `tone') is used to symbolise 9/8
and `S'  (for `semi-tone') to  symbolise 256/243 the  pattern obtained
would be  - T  T S T  T T  S...  This  is the same  as the  pattern of
intervals between the  white notes on a  piano - a T means  there is a
black note between the  two white notes and S means  there is no black
note in between. This pattern defines  the scales as well as the notes
in the  western tradition.   To obtain  a scale, a  base or  home note
needs to be specified. Hence, up to octave equivalence there are seven
notes to choose from, giving rise to the following seven patterns -
\ber
&& \mbox{T T S T T T S \; \; [C] \; \; (major)} \nonumber \\
&& \mbox{T S T T T S T \; \; [D]} \nonumber \\
&& \mbox{S T T T S T T \; \; [E]} \nonumber \\
&& \mbox{T T T S T T S \; \; [F]} \nonumber \\
&& \mbox{T T S T T S T \; \; [G]} \nonumber \\
&& \mbox{T S T T S T T \; \; [A] \; \; (minor)} \nonumber \\
&& \mbox{S T T S T T T \; \; [B]} \nonumber 
\eer
The first  of these is obtained  if one starts  at a “C” on  the piano
keyboard. Historically, all  of these scales were used,  but over time
only   the  first   (major  scale)   and  the   sixth  (minor   scale)
survived\footnote{It should be noted that the Indian {\em saptak} is
  actually a {\bf \em major} scale,  allowing for the base note to
  have any frequency at all.}.

Even though this method of setting  up a musical scale has mostly been
attributed to  Pythagoras, there have  also been others  (for example,
Eratosthenes)  responsible  for  developing  the  scale.   Pythagorean
tuning  is well  suited to  music that  emphasises musical  fifths and
octaves and  has been  used almost  exclusively by  European musicians
till  about  16th century.  Unfortunately,  intervals  other than  the
octave  and the  fifth  are  not quite  perfect  in  this tuning.   In
particular, the major third (81/64) and minor third (32/27) are rather
dissonant. Moreover, there is very little freedom to modulate from the
home key  to more distant  keys. Because, going  up a fifth  a certain
number of times never equals going up  by a number of octaves (see the
discussion in  \cref{s-etscl}). As musical thirds  (intervals defined
by  frequency ratios  of 6/5,  5/4) became  more important  to musical
expressions, and as  the desire to move far beyond  the home key grew,
other alternative temperaments were developed later on.

\section{Equal Tempered Scale}
\label{s-etscl}
It has been seen that the Pythagorean scale has basically been generated
by taking a base note and multiplying it by 3/2 again and again, modulo
a factor of 2 (to account for the octave equivalence). It is to be noted
that -
\beq
(3/2)^{12} = 129.746 \simeq 1.013 \times 2^7 \,.
\eeq
In other words, it takes 12 steps around what is called the `circle of
fifths'  until an  approximate power  of  two (an  integral number  of
octaves  above)  of the  original  note  is reached.   Therefore,  the
simplest explanation for the twelve tones - C, C\#, D, D\#, E, F, F\#,
G, G\#, A, A\#, B (\# denotes a  sharp note) - within an octave is the
above relation between 3/2 and 2.  However, as $(3/2)^{12}/2^7$ is not
exactly equal to  1, powers of 3/2 (or `circles  of fifths') would not
take us exactly to twice the original frequency~\footnote{Because, the
  `circle of fifth' does not quite close in Pythagorean tuning, one of
  the intervals must  {\em NOT} match the  prescribed frequency ratio,
  in  order to  close  the circle  forcibly.  Then  there  would be  a
  dissonant beat.   This is known  as a {\em  `wolf interval'}(meaning
  the interval howls like a wolf!).}

In an  attempt to  maximise the number  of consonant  intervals having
exact frequency ratios  within the octave the {\em  `Just Scale'} (also
known   as  the   `harmonic   tuning'  or   `Helmholtz's  scale')   was
created. This scale starts from a perfect triad - a base note ($\nu$),
the perfect  third (4$\nu$/3) and  a perfect fifth (3$\nu$/2)  - which
has a  frequency ratio of 4:5:6.  The construction of the  rest of the
scale is  as before. One goes  up from a  given note by a  major third
($\nu \rightarrow  4\nu$/3) or  by a perfect fifth  ($\nu \rightarrow
3\nu/2$) and the resulting note  is brought back to the home octave by
descending  an   appropriate  number  of  octaves   ($\nu  \rightarrow
\nu/2^n$) till the note having twice (or approximately twice) the base
note is achieved.

However, the  Just Scale  too suffers from  problems similar  to those
encountered by the  Pythagorean scale. It appears to  be impossible to
create  a  tuning system  that  has  perfect  intonation for  all  the
consonant musical intervals within the octave. There appears to be two
basic problems - \\
  {\bf a)} to  have perfect intonation  (correct ratio of  frequencies)
  for all consonant musical intervals within the octave, \\
  {\bf b)} to have  freedom to modulate, to move away  from the base
  octave without losing consonance. 

The ETS, created around the 19th century, specifically addresses these
particular problems.  Since there are  12 notes the octave  is divided
equally (in a logarithmic sense) in 12 steps and each note is obtained
by   multiplying  the   previous   one   by  $2^{\frac{1}{12}}   (\sim
1.0595)$. The frequency ratios from the base note in this scale are as
follows  - 1.0000,  1.0595,  1.1225, 1.1892,  1.2599, 1.3348,  1.4142,
1.4983,  1.5874,  1.6818,  1.7818,  1.8877,  2.0000.   Evidently,  all
octaves would  be perfect  now, with ratio  2:1.  Therefore,  there is
absolute freedom to modulate from one octave to another as there is no
distinction  between  them  –  they  all  contain  the  same  semitone
intervals,  and have  the  same  values for  all  the other  (slightly
compromised) intervals as well.  Thus, the actual frequencies obtained
are  slightly  different from  the  Pythagorean  scale.

\begin{table}[h]
\vspace{-0.25cm}
\begin{tabular}{llll} 
Notes (major scale) & C4 & E4 & G4 \\
Pythagorean (Hz) & 260.741 & 330.00 & 391.11 \\
ETS (Hz) & 261.625 & 329.63 & 392.00 \\
\end{tabular}
\vspace{-0.5cm}
\end{table}
However,  the  difference between  the  frequencies  obtained using  a
Pythagorean  scale   and  the   ones  obtained   in  ETS   are  rather
small. Assuming  $A$ to  be at  440~Hz (concert  tuning)\footnote{ The
  absolute pitch, or base key, in  the baroque period used to be tuned
  at  A =  415~Hz, which  was  used by  Bach while  Handel used  422.5
  Hz. The  international tuning  standard of  A4 =  440 Hz  was widely
  adopted around 1920.}, the frequencies of the notes of a major chord
are as seen  above.  It is evident that the  differences are too small
to be perceptible  to the human ear.   This is the reason  why, at the
present  time, the  keys  of  most reed  instruments  (the piano,  the
electronic keyboard  or the harmonium\footnote{Harmonium  - Originally
  an European reed  instrument, a modified pump  organ, was introduced
  to Asia by Christian missionaries. It gained huge popularity in Asia
  as an  accompaniment and was  modified to suit the  Asian musician's
  practice of sitting on the floor.}) are tuned according to the ETS.

There is an interesting way to  see the connection between the ETS and
the  consonance of  fifth.  Basically,  ETS is  being  defined by  the
relation  $(3/2)^m  \simeq 2^n$  where  both  $m,n$ are  integers  (Of
course, the  exact equality can  never hold true.). Therefore,  we are
trying to represent a real number $x$ by the quotient of two (not very
large) integers $m$ and $n$ where $x$ is given by -
\beq
3/2 = 2^x
\; \Rightarrow \; x = \frac{\log{3/2}}{\log 2} = 0.584962500721... 
\eeq
Now, any real number can be expressed in terms of a continued fraction
where the expansion has 1 in all the numerators and continues forever.
The reason for  using continued fractions is that these  give the best
rational   approximations   to   real  numbers,   i.e.    any   closer
approximation would have a larger  denominator.  When $x$ is expressed
thus, we obtain -
\beq
\Large{
\frac{\log{3/2}}{\log 2} =
  \frac{1}{1 + \frac{1}{1 + \frac{1}{2 + \frac{1}{2 + \frac{1}{3 + \frac{1}{1 +
              \frac{1}{5 + \frac{1}{2 + \frac{1}{23 + \frac{1}{2 + ..}}}}}}}}}}
}
\eeq
Taking the first few terms as approximate values for this continued fraction
we obtain -
\ber
&& \mbox{Approximation 1.} \; \; \; \; \; \; \frac{1}{1 + \frac{1}{1}} = \frac{1}{2} = 0.5 \\
&& \mbox{Approximation 2.} \; \; \; \; \; \; \frac{1}{1 + \frac{1}{1 + \frac{1}{2}}} = \frac{3}{5} = 0.6 \\
&& \mbox{Approximation 3.} \; \; \; \; \; \; \frac{1}{1 + \frac{1}{1 + \frac{1}{2 + \frac{1}{2}}}} = \frac{7}{12} = 0.58\dot{3} \\
&& \mbox{Approximation 4.} \; \; \; \; \; \; \frac{1}{1 + \frac{1}{1 + \frac{1}{2 + \frac{1}{2 + \frac{1}{3}}}}} = \frac{24}{41} = 0.5854..\,... 
\eer
The approximation  improves when  more number  of terms  are included.
However, when more terms included the fraction is obtained in terms of
larger and larger  integers.  Evidently, the ratio 7/12  appears to be
an excellent  compromise which  immediately suggests  an octave  of 12
steps.  Moreover,   it  should   be  noted  that   $2^{\frac{7}{12}}  =
1.498.. \simeq  3/2$ which shows  that a  fifth of the  Pythagorean is
equal to 7 steps of this equal tempered scale.

With time ETS  has all but replaced the original  Pythagorean scale. In
particular, most reed instruments are  tuned to this scale. Of course,
musicians  using  string  instruments  and vocalists  still  have  the
freedom to choose  a scale according to their convenience.  One of the
traditions where this freedom has  been of paramount importance is the
Indian classical music.  In the next and final article  of this series
we shall discuss the scales used in that genre.

\end{document}